\documentstyle[aps,psfrag,graphicx,preprint,tighten,eqsecnum,floats,epsf,epsfig,rotate,prd,here]{revtex}
\newcommand{\be}{\begin{eqnarray}}
\newcommand{\ee}{\end{eqnarray}}
\newcommand{\sbe}{\begin{eqnarray*}}
\newcommand{\see}{\end{eqnarray*}}
\newcommand{\pslash}{\not\!p}
\begin{document}
\draft
\vskip 4mm\preprint{
Preprint No. ADP-00-51/T431}

%_______________________ Title, Authors ____________________________________

\title{Regularization-independent studies of nonperturbative field theory}

\author{
        Ay{\c s}e K{\i}z{\i}lers{\" u}
            \footnote{E-mail:~akiziler@physics.adelaide.edu.au},
        Andreas W.  Schreiber
            \footnote{E-mail:~aschreib@physics.adelaide.edu.au}
          and
        Anthony G.\ Williams
            \footnote{E-mail:~awilliam@physics.adelaide.edu.au}
	\vspace*{5mm}
        }

\address{
   Special Research Centre for the Subatomic Structure of Matter and \\
   Department of Physics and Mathematical Physics, \\
   Adelaide University, 5005, Australia
   \vspace*{2mm}
        }

%\date{}
%
%%-------------------------------------------------------------------%
%

\maketitle

\begin{abstract}

We propose a regularization-independent method for studying a
renormalizable field theory nonperturbatively through its
Dyson-Schwinger equations.  Using QED$_4$ as an example, we show how
the coupled equations determining the nonperturbative fermion and
photon propagators can be written entirely in terms of renormalized
quantities, which renders the equations manifestly finite in a
regularization-independent manner.  As an illustration of the
technique, we apply it to a study of the fermion propagator in
quenched QED$_4$ with the Curtis-Pennington electron-photon vertex.
At large momenta the mass function, and hence the anomalous mass
dimension $\gamma_m(\alpha)$, is calculated analytically and we find
excellent agreement with previous work.  Finally, we show that for the
CP vertex the perturbation expansion of $\gamma_m(\alpha)$ has a
finite radius of convergence.

\end{abstract}

\section{Introduction}
\label{sec_intro}

As is well known, in studies of renormalizable relativistic field
theories it is unavoidable to make use of a regulator in order to
control the UV divergences of the theory, even though at the end this
regulator is removed in order to obtain physical results.  In
perturbation theory the regulator and its removal via the
renormalization procedure is usually little more than an inconvenience,
however in nonperturbative studies, which are largely numerical, much
effort needs to made in order to perform this removal in a
well-controlled manner.  

The best known example of the latter, of course, is lattice gauge
theory\cite{lattice_QED} where the regulator is the lattice spacing
and its removal entails a careful numerical extrapolation to the
continuum.  Another popular tool, which is the subject of this
investigation, is the numerical study of a truncated set of the
Dyson-Schwinger equations of the
theory~\cite{FGMS_Review,Miransky_book,RW_Review}, where most often
the regulator is simply a hard momentum cutoff $\Lambda$.  Although
these calculations are numerically far less demanding than the
numerical integration of a functional integral, they are not entirely
trivial either as they involve solving a set of integral equations for
momenta $p$ ranging over many orders of magnitude, from $|p|=0$ to
$|p|=\Lambda$.  Renormalization then requires a further numerical
extrapolation, with fixed boundary conditions provided by the relevant
renormalization scheme, to $\Lambda \rightarrow \infty$.

That this removal of the regulator can be done numerically was
demonstrated explicitly in
Refs.\cite{qed4_hw_etal0,qed4_hw_etal1,qed4_hw_etal2} within
subtractively renormalized quenched QED, using an off-shell
renormalization scheme.  However, as emphasized in those works and
pointed out first by Dong, Munczek and Roberts~\cite{DMR} (it was also
apparently realized at an earlier stage by Pennington~\cite{mike1}),
the use of a momentum cut-off as a regulator is not entirely
satisfactory as it violates translational invariance (in momentum
space) which leads to a violation of gauge invariance in the final
results, {\em even after} $\Lambda$ is taken to infinity.  More
precisely, this violation of gauge invariance was observed in quenched
QED calculations employing the Curtis-Pennington (CP) electron-photon
vertex and may be traced back to a certain, logarithmically divergent,
4-dimensional momentum integral which vanishes because of rotational
symmetry at all $\Lambda < \infty$, but leads to a finite contribution
for $\Lambda \rightarrow \infty$.  It is this discontinuous behaviour
as a function of $\Lambda$ which complicates correct numerical
renormalization with this regulator.  Incorrect results will be
obtained unless care is taken to drop `gauge covariance violating
terms'.  Although this recipe appears to work in practise (see the
Appendix of Ref.~\cite{qed4_hw_etal1} as well as Ref.~\cite{KSW_regn}
for a more detailed discussion) it is fair to say that it is ad hoc
and it would be more desirable to avoid it altogether.

Dimensional regularization, of course, does not break translational
invariance and hence will not lead to the spurious gauge covariance
violations outlined above.  However, as shown in
Ref.~\cite{qed_dim_reg}, the use of this regulator in nonperturbative
studies of chiral symmetry breaking introduces a new unwelcome
feature: In a dimensionally regulated theory a fermion mass will be
generated dynamically at all values of the coupling constant $\alpha$
in $D \ne 4$ dimensions for any vertex which only does so above a
finite critical coupling $\alpha_{cr}$ in $D=4$ dimensions.  In other
words, again discontinuous behaviour (this time as a function of
$4-D$) prevents a naive removal of the regulator.  Ultimately, it is
this feature which limited the accuracy with which one could determine
the critical coupling associated with the CP vertex within
dimensional regularization in Ref.~\cite{qed_dim_reg}.  In addition,
of course, with this regulator the integration range now extends to
$|p| \rightarrow \infty$, compounding the numerical difficulties
mentioned at the start.  In short, although it was shown in
Ref.~\cite{qed_dim_reg} that the results obtained within dimensional
regularization agree with those obtained by employing a cut-off
within some limited precision, the numerical difficulties encountered
in implementing this regulator has limited its further use.

It is the purpose of this letter to point out that all the difficulties
mentioned above may be circumvented by removing the regulator {\em
analytically} rather than {\em numerically}. This may be done by
formulating the Dyson-Schwinger equations in terms of renormalized
quantities only, where the dependence on the mass scale introduced by
the regulator is traded for the momentum scale $\mu$ at which the
theory is renormalized {\em before} performing any numerical
calculations.  This way dimensional regularization (or any other
regulator which doesn't violate gauge covariance) can be used and,
more importantly from a numerical point of view, removal of the
regulator means that no longer does one have to solve integral
equations involving mass scales of vastly different orders of
magnitude (and then, in addition, take a limit in which one of the
scales goes to infinity).  Rather, the important scales in the problem
become scales of `physical' importance, such as the renormalized mass
$M_\mu$ and the renormalization scale $\mu$ at which this mass is
defined.  It is therefore to be expected that the dominant
contributions to any integrands will be from a relatively small region
of momenta.  This feature is generic; i.e. it is independent of the
particular vertex Ansatz that one makes use of and it remains a valid
consideration for an arbitrary renormalizable field theory.  Hence, although we shall
illustrate this approach only within QED (and numerically only
within quenched QED), it is our hope that ultimately this
`regularization-free' approach will prove its utility elsewhere.

To a certain extent there is a price to be paid for using the approach
advocated above: one looses all contact to the bare theory. In
particular, we stress that because of this one cannot study dynamical
chiral symmetry breaking in this approach by simply setting the bare
mass $m_0$ equal to zero and investigating at what point a dynamical
fermion mass is generated.  It is in this crucial point that we differ
from the analysis in Ref.~\cite{CP92} where, within quenched QED, a
removal of the regulator was attempted in a manner which has some
similarity to what we do here.  Indeed, as emphasized particularly by
Miranksy (Refs.~\cite{noreg},~\cite{FGMS_Review} as well as Ch 10.7 of
Ref.~\cite{Miransky_book}), some care needs to be taken with the
treatment of the bare mass while removing the regulator in order to
avoid drawing incorrect conclusions about the presence or absence of dynamical
chiral symmetry breaking in gauge theories.  In the present approach
this problem is avoided: chiral symmetry breaking, both explicit as
well as dynamical, is characterized by a nonzero renormalized mass
$M_\mu$, and hence this quantity in itself does not distinguish
between these two possibilities.  In Section III, we shall take (within
quenched QED) the somewhat more indirect signal of the
appearance of oscillations in the renormalized fermion mass 
function~\cite{noreg}
as an indicator of the onset of dynamical chiral symmetry breaking.

This paper is organized as follows: in the next section all reference
to the bare Lagrangian will be explicitly eliminated by re-expressing
the Dyson-Schwinger equations of full QED in terms of renormalized
quantities only.  As a first application of this seemingly innocuous
step we then re-examine, in the second part of this paper, the large
momentum behaviour of the electron propagator in quenched QED$_4$
(employing the Curtis-Pennington vertex~\cite{mike2}).  In particular,
because of the absence of a cut-off, scaling invariance in this
momentum region is restored in the equation for the electron's mass
function and hence this equation can be solved analytically.  We compare the
results with existing numerical studies and also use it to derive the
analytic form of the anomalous mass dimension $\gamma_m(\alpha)$ within this offshell
momentum subtraction scheme.  Finally, we discuss the analytic behaviour
of $\gamma_m(\alpha)$ as a function of the coupling and conclude.

\section{Regularization-independent formulation}
\label{sec_reg_indep}

\begin{figure}[htp]
\begin{center}
~\epsfig{file=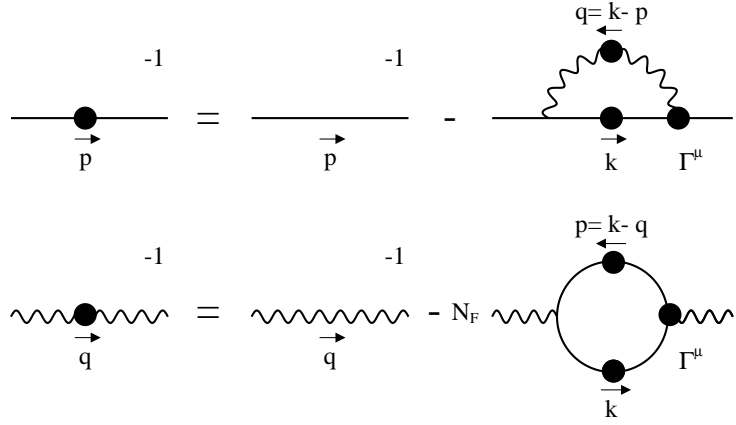,width=300pt}
\end{center}
\caption{The Dyson-Schwinger equations for the fermion and photon
propagators in QED}
\label{fig:vertex}
%\vspace{-1cm}
\end{figure}

The Dyson-Schwinger equations for the renormalized
electron and photon propagators
(Fig.~\ref{fig:vertex}) are well known.  For example, employing
dimensional regularization, we have
\be
S^{-1}(\mu;p) &=& Z_2(\mu) {S^{0}}^{-1}(p)
\> - \> i Z_1(\mu) {e_\mu}^2 \int {d^D k \over (2 \pi)^D}\>
\Gamma^\alpha (\mu;p,k) \>  S(\mu;k) \gamma^\beta \> 
 D_{\alpha \beta}(\mu;q)
\label{eq:mainsdf} \\
&\equiv& Z_2(\mu) {S^{0}}^{-1}(p)
\> - \>  Z_1(\mu) \> \bar \Sigma(p)\nonumber \\
D_{\alpha \beta}^{-1}(\mu;q)  &=& Z_3(\mu) {D_{\alpha \beta}^{0}}^{-1}(q)
\> + \>i  Z_1(\mu) {e_\mu}^2 N_F \> {\rm tr}  \int {d^D k \over (2 \pi)^D}\>
\Gamma_\alpha (\mu;p,k)\>S(\mu;k) \>\gamma_\beta \>  S(\mu;p)\>
\label{eq:mainsdph} \\
&\equiv& Z_3(\mu) {D_{\alpha \beta}^{0}}^{-1}(q)
\> + \>  Z_1(\mu) \> \bar \Pi_{\alpha \beta}(q)\nonumber
\ee

Here $N_F$ is the number of species of fermions in the theory and
the constants
$ Z_1(\mu)$, $ Z_2(\mu)$ and $ Z_3(\mu)$ are the vertex, 
fermion wave function and photon wave function renormalization
constants, respectively. We have assumed, for convenience,
 that the $N_f$ fermions
are identical (i.e. they have equal mass), the generalization being
straightforward. $S(\mu;p)$  and $D_{\alpha \beta}(\mu;q)$ 
 are the full fermion and photon
propagators  renormalized at the momentum scale $\mu$ ($S^{0}(p)$
and $D_{\alpha \beta}^{0}(q)$ are their bare counterparts), while
the full (proper) fermion-photon vertex is 
$\Gamma_{\alpha}(\mu;k,p)$.
These propagators
are usually expressed in terms of scalar functions $Z(\mu^2;p^2)$,
$M(p^2)$ and $G(\mu^2;q^2)$ through
\be
S(\mu;p) \>=\> \frac{Z(\mu^2;p^2)}{\not\!p - M(p^2)}
\hspace{15mm}
D_{\alpha\beta}(\mu;q)\>=\>-\frac{1}{q^2}\,
\left[G(\mu^2;q^2)
\left(g_{\alpha\beta}-\frac{q_{\alpha}q_{\beta}}{q^2}\right)
+\xi\frac{q_{\alpha}q_{\beta}}{q^2}
\right],
\ee  
where $\xi$  is the renormalized  covariant gauge 
fixing parameter, 
so that the fermion's equation separates into two scalar equations
\be
Z^{-1}(\mu^2;p^2) &=&  Z_2(\mu) \>-\> Z_1(\mu) \>\bar \Sigma_d(p^2)
\label{eq: z eq}\\
M(p^2)\> Z^{-1}(\mu^2;p^2)&=&  Z_2(\mu)\> m_0\> +\> Z_1(\mu)\>\bar \Sigma_s(p^2)\;\;\;.
\label{eq: m eq}
\ee
Here $\bar
\Sigma_d(p^2)$ and $\bar \Sigma_s(p^2)$  refer,  respectively,
to the Dirac odd and even parts of $\bar \Sigma(p)$, i.e. 
$\bar \Sigma(p)= \pslash \;\bar\Sigma_d(p^2) + \bar\Sigma_s(p^2)$.  The bar 
over these quantities
indicates that we have explicitly separated out the renormalization
constants $Z_i(\mu)$;  also, note that we do not indicate the
implicit dependence of $\bar \Sigma_{d,s}(p^2)$,
through the functions $Z(\mu^2;p^2)$  and
$D_{\alpha\beta}(\mu;q)$, 
on $\mu^2$.  The equivalent equation 
for the photon wavefunction renormalization $G(\mu^2;q^2)$ is given by
\begin{equation}
G^{-1} (\mu^2;q^2) \> = \>  Z_3(\mu) \> + \> Z_1(\mu) \>\bar \Pi(q^2)
\;\;\;,
\label{eq: g eq}
\end{equation}
where $\bar \Pi^{\alpha \beta}(q) \equiv - q^2 \left(g^{\alpha \beta}-\frac{q^{\alpha}q^{\beta}}{q^2}\right)\bar \Pi(q^2)$.
Note that, as we are free to use a regulator which respects gauge covariance
and will use a vertex which satisfies the Ward-Takahashi identity,
$\bar \Pi^{\alpha \beta}(q)$ does not
contain spurious quadratic divergences~\cite{quaddiv} and
is transverse as it should be. Hence
the term involving the gauge parameter $\xi$ on the left hand
side of Eq.~(\ref{eq:mainsdph}) precisely cancels the corresponding term
involving $\xi_0$ in $D_{\alpha \beta}^0$ on the right hand side
(i.e. $\xi_0=Z_3 \xi$).  

The renormalization constants $Z_i(\mu)$ as well as $m_0$ are fixed by the 
boundary conditions
\be
Z(\mu^2;\mu^2) \>=\> 1 \quad\quad\quad
M(\mu^2) \>\equiv \> M_\mu \quad\quad\quad
G(\mu^2;\mu^2) \>=\> 1 \label{eq:renorm cond}\;\;\;.
\ee
(Although we shall not do so here, in practice the fermion and photon can 
be renormalized at two
different scales $\mu^2$ and ${\mu'}^2$.  In QED it is usual
to choose the latter as ${\mu'}^2=0$, in which case ${e_{\mu'}}^2/(4 \pi)$
just becomes the physically measured fine structure constant $\alpha$.)
Because of the Ward-Takahashi identity, which should be satisfied
by the vertex Ansatz for $\Gamma_\alpha$,  one has $ Z_1(\mu)= Z_2(\mu)$.
 
In order to avoid cumbersome notation, we have not explicitly
indicated functional dependence on the regulator in
Eqs.~(\ref{eq:mainsdf})--(\ref{eq: g eq}).  Most of the above
quantities, i.e.  $Z(\mu^2;p^2)$, $M(p^2)$ and $G(\mu^2;q^2)$, as
well as $Z_{1,2,3}(\mu)$, are functions of the regulator.  Indeed, if one
keeps $M_\mu$ and $e_\mu$ fixed as one removes the regulator, which is
the standard textbook prescription,  the
bare fermion mass $m_0$ and the bare charge also become regulator dependent.

As one removes the regulator, the integrals on the right hand side of
Eqs.~(\ref{eq:mainsdf}) and~(\ref{eq:mainsdph}), and hence
$\bar \Sigma_d(p^2)$, $\bar \Sigma_s(p^2)$ and $\bar \Pi(q^2)$
diverge logarithmically (or, to be more precise, in dimensional
regularization they develop singularities at $D=4$).  It is the
defining feature of a renormalizable field theory that these
divergences may be absorbed into the constants $Z_{1,2,3}(\mu)$ and into
the bare mass $m_0$, rendering finite limits for $Z(\mu^2;p^2)$,
$M(p^2)$ and $G(\mu^2;q^2)$.  However, we can make use of the $p^2$
independence of $Z_{1,2,3}(\mu)$ and $m_0$ in order to eliminate these 
constants
 from
the above equations. For example, by evaluating Eq.~(\ref{eq: z eq})
at a second momentum which, for convenience, we take to be $p^2=\mu^2$, 
and forming an appropriate difference one obtains 

\be
Z(\mu^2;p^2) \> = \> 1 \> + \> Z(\mu^2;p^2) \>\bar \Sigma_d(p^2)
\> - \> \bar \Sigma_d(\mu^2)
\label{eq:subwavefunc}
\ee
where we have made use of the renormalization condition~(\ref{eq:renorm cond}).
  Because the left hand side of this equation
is finite as the regulator is removed, the right hand side must
also be (even though, of course, the individual terms on the RHS separately
diverge).  In a similar way one obtains
\be
M(p^2) \> = \> M_\mu \> + \> \left [M(p^2) \bar \Sigma_d(p^2)
\> + \>\bar \Sigma_s(p^2)\right]
\> -  \> \left[M_\mu \bar \Sigma_d(\mu^2) \> + \> \bar \Sigma_s(\mu^2)\right]
\label{eq:submass}
\ee
and
\be
G^{-1}(\mu^2;q^2)\> = \>1 \> + \>
\left[ G^{-1}(\mu^2;q^2)\bar \Sigma_d(\mu^2) \> + \>
\bar \Pi(q^2)\right] \> - \>\left[ \bar \Sigma_d(\mu^2)\> + \>
\bar \Pi(\mu^2) \right]
\;\;\;.
\label{eq:subphoton}
\ee

We note in passing that the perturbative solution of these three equations
defines `finite QED' in a rather different sense to that developed by
Johnson, Baker and Willey~\cite{JBW}.  We do not pursue this further here; 
rather, the main point of this letter is to point out that these renormalized
equations~(\ref{eq:subwavefunc}) to~(\ref{eq:subphoton}) (with the
regulator removed, of course), or equivalently their counterparts in
QCD or any other renormalizable field theory, provide a starting point
for nonperturbative investigations which has some significant advantages
over the usual treatment found in the literature.  To illustrate this
we now turn to the well-studied example of quenched QED.

\section{Application to Quenched QED$_4$}

For quenched QED Eq.~(\ref{eq:subphoton}) becomes irrelevant
and Eqs.~(\ref{eq:subwavefunc}) and~(\ref{eq:submass}) reduce to
one-dimensional integral equations.  In particular, if
we employ the CP vertex (which, importantly for the present approach,
satisfies the Ward-Takahashi identity and respects renormalizability)
and make use of a regulator which does not violate gauge
covariance then these renormalized equations become, in Euclidean
space,
\be
Z(\mu^2;p^2)  & = &  1
-\frac{\alpha\xi}{4\,\pi}\,\int_{p^2}^{{\mu}^2}
dk^2 \frac{1}{\left[k^2+M^2(k^2)\right]}\,Z(\mu^2;k^2)\nonumber\\
&& \hspace{1.5cm}
+\frac{\alpha}{4\,\pi}\,\int_0^{\infty} 
		\frac{dk^2}{\left[k^2+M^2(k^2)\right]}
    \left [Z(\mu^2;p^2)\,I(k^2,p^2) \> - \> I(k^2,\mu^2) \right ]
\label{eq:renwavefunc}\\
M(p^2)  & = & M_\mu
+\frac{\alpha}{4\pi}\int_0^{\infty}\frac{dk^2}{\left[k^2+M^2(k^2)\right]}
\left [ J(k^2,p^2) - J(k^2,\mu^2) 
\right.\nonumber \\&&\left.\hspace{5cm}
\> + \> M(p^2)\>I(k^2,p^2) - M_\mu \>
I(k^2,\mu^2)\right ]
\label{eq:renmass}
\ee
where $I(k^2,p^2)$ and $J(k^2,p^2)$ are those of Table.~1 of  Atkinson 
et al.'s paper~\cite{Atkinson}, apart 
from the term proportional to $\xi\left (1+{M^2(k^2) \over k^2}\right)$ 
in $I(k^2,p^2)$ in their paper which
originated from the violation of gauge covariance by their cutoff mentioned
in the introduction.
As promised, these equations are finite in the absence of a
regulator.  The equation for the wavefunction renormalization
function (Eq.~\ref{eq:renwavefunc}) is the same as Eq. (10) in Ref.~\cite{CP92},
however our treatment of the mass function differs from theirs.
We shall not solve these equations numerically here but, rather,
shall make use of the fact that, for momenta $k^2$ larger than
$M^2(k^2)$, the absence of a cut-off implies that they become
scaling invariant.  Hence we can solve them analytically in this
region for finite $M_\mu$, in the same way as was done by Atkinson et 
al.~\cite{Atkinson} for the chirally symmetric theory (i.e. $M_0=M_\mu=0$).

\subsection{Large Momentum Behaviour in Quenched QED$_4$}
For $k^2 >> M^2(k^2)$ Eqs.~(\ref{eq:renwavefunc},~\ref{eq:renmass})
may be expanded in powers of ${M^2(k^2) \over k^2}$.  Keeping at most
linear terms in $M(k^2)$ we obtain

\be
Z(\mu^2;p^2)&=&1\>-\>\frac{\alpha \xi}{4\pi} \int_{p^2}^{\mu^2} 
\frac{dk^2}{k^2} Z(\mu^2;k^2)\quad,  
\label{eq:bifwavefunc}  \\    
M(p^2)&=&M_\mu \>+ \> {\alpha \xi \over 4 \pi}
\left \{ \int_0^{p^2} {d k^2 \over p^2} {Z(\mu^2;k^2) \over   Z(\mu^2;p^2) }
M(k^2) \> - \> \int_0^{\mu^2} {d k^2 \over \mu^2} Z(\mu^2;k^2) M(k^2) \right \}
\nonumber \\
&&\>+\> {3 \alpha  \over 8 \pi}
 \int_0^\infty d k^2 \left \{
\left [ 2 \, M(k^2) {p^2 \over p_>^2}
{ {Z(\mu^2;k^2) \over   Z(\mu^2;p^2) } - {k^2 \over p^2} \over  p^2-k^2}
\right.\right.\nonumber \\
&&\left.\left.\hspace{15mm}
\> - \> {Z(\mu^2;k^2) \over   Z(\mu^2;p^2) }
{p_<^2 \over p_>^2} { M(p^2) - M(k^2) \over  p^2-k^2} \right ]
\> - \>  [p^2 \rightarrow \mu^2 ]\right \}\;\;,
\label{eq:bifmass}
\ee
where $p_<^2$ ($p_>^2$) is the minimum (maximum) of $p^2$ and $k^2$.

Because of scaling invariance, these equations are solved by a
power law Ansatz for $Z(\mu^2;p^2)$ and $M(p^2)$, i.e.
\be
Z(\mu^2;p^2) \>=\>\left(\frac{p^2}{\mu^2}\right)^{\nu} \quad\quad
M(p^2)  \>=\>  M_\mu \left(\frac{p^2}{\mu^2}\right)^{-s}\;\;\;.
\ee
The reader should note that the overall scale of $M(p^2)$ is fixed
by the renormalization condition.
Explicitly, one finds $\nu$ and $s$ to be given by
\be
\nu=\frac{\alpha \xi}{4\pi}\quad, 
\label{eq:nu}
\ee
\be
&&\frac{3\nu}{2\xi}\Bigg[
		2\pi\cot{s\pi}-\pi\cot{\nu \pi}+3\pi\cot(\nu-s)\pi
		+\frac{2}{(1-s)}+\frac{1}{(\nu+1)}+\frac{1}{\nu}\nonumber\\
&&\hspace{6cm}-\frac{3}{(\nu-s)}-\frac{1}{(\nu-s+1)}\Bigg]
	        +\frac{s-1}{\nu-s+1}
		= 0 \quad. 
\label{eq: s def}
\ee

\begin{figure}[H]
\centerline{
%   \parbox{6cm}{ 
	\epsfig{figure=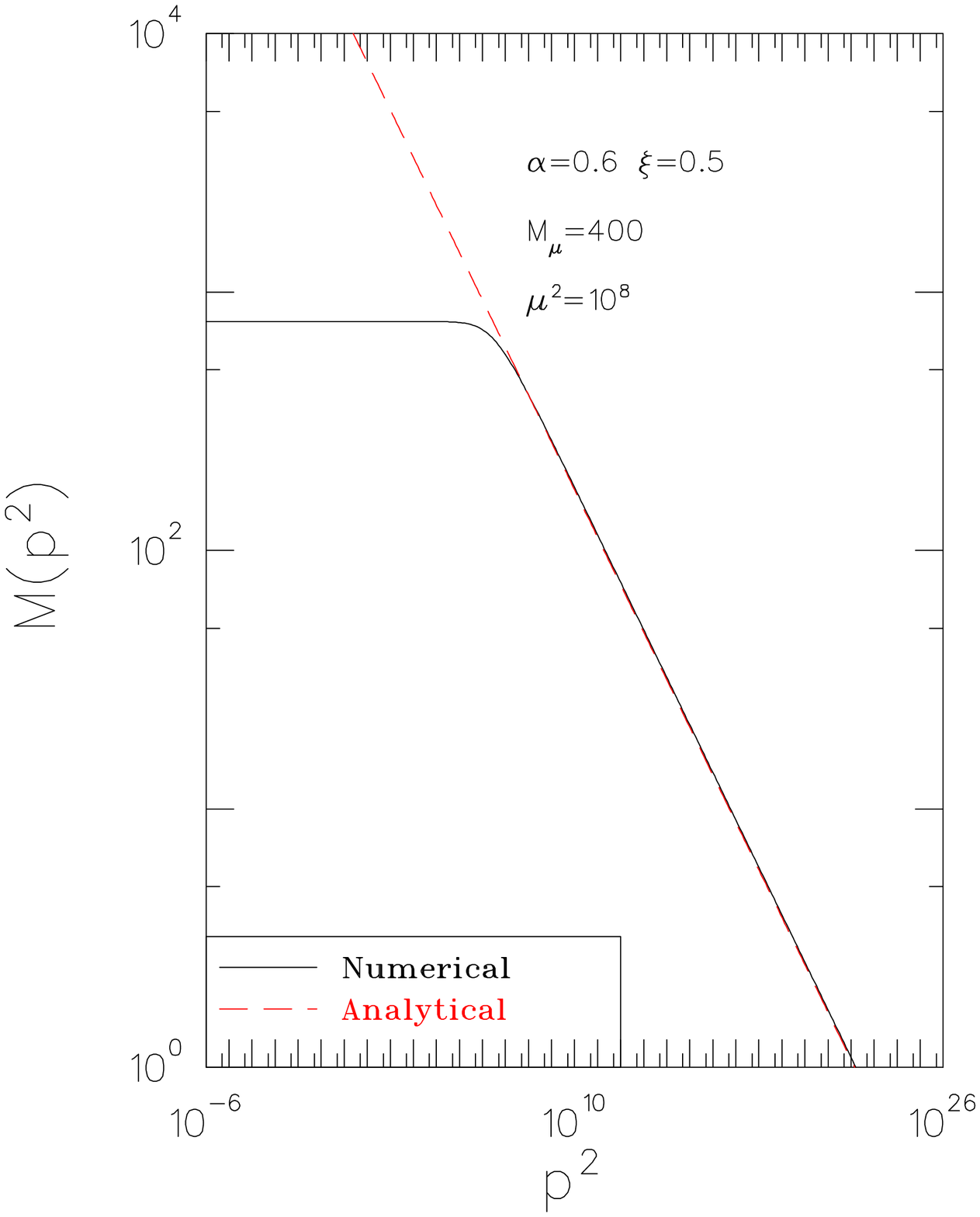,width=7cm} 
%}
	\hspace {1cm}
%   \parbox{10cm}{ 
	\epsfig{figure=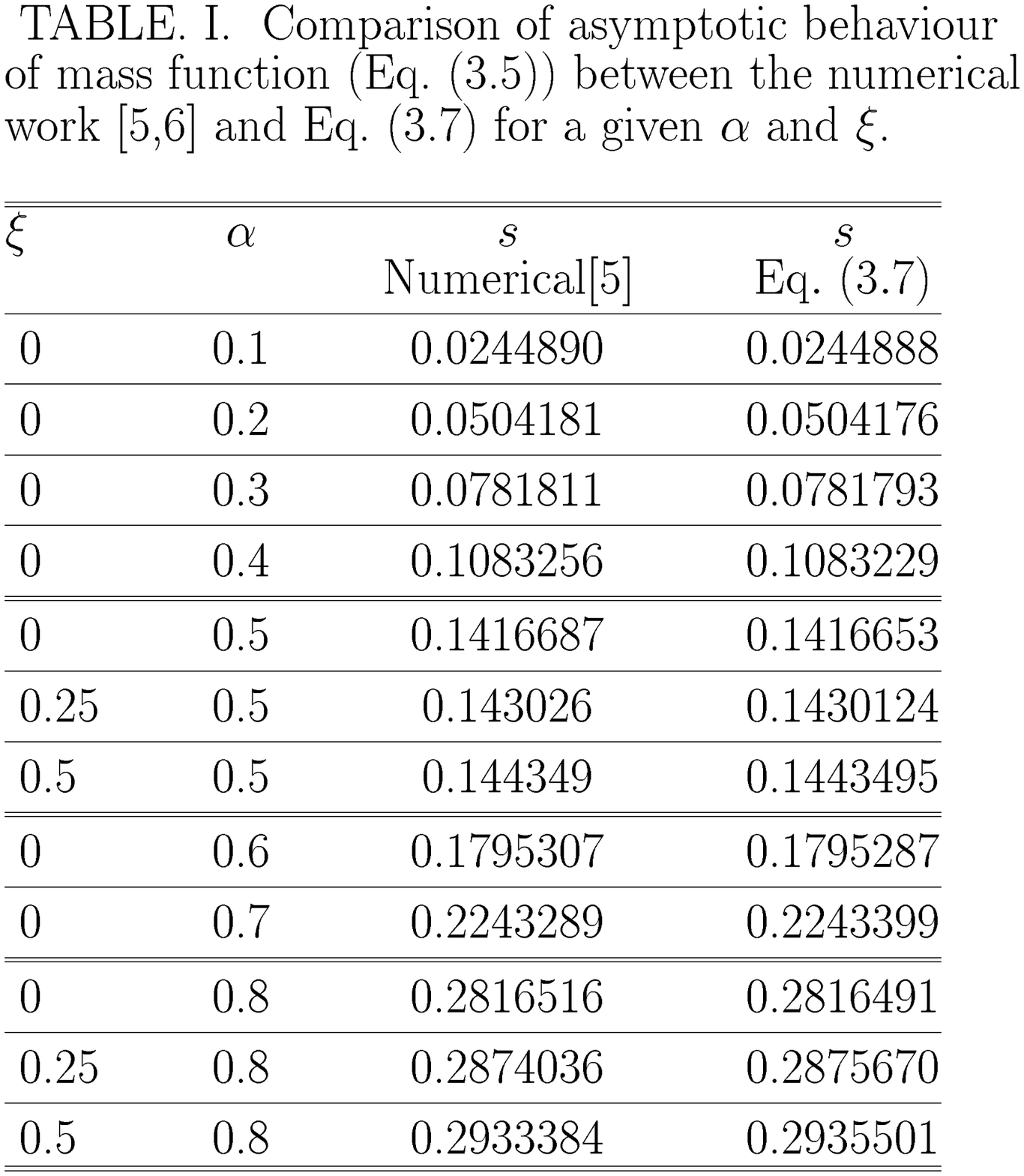,width=7cm} 
%}
           }
%
%\hspace*{0.5cm}
%\refstepcounter{figure}
%\addtocounter{figure}{-1}
%~\epsfig{file=fig1al08xi05.ps,width=7cm}
%\end{center}
\caption{Comparison of a numerical solution for the mass function $M(p^2)$
(solid line,
obtained from Ref.~\protect\cite{qed4_hw_etal1})
and the analytical expectation for its asymptotic behaviour
(dashed line, Eq.~\protect\ref{eq: s def}). The renormalized mass is 
$M_\mu=400$ (arbitrary units), while $\alpha=0.6$ and $\xi=0.5$. }
\label{fig:numan}
%\vspace{-1cm}
\end{figure}

These are the same powers as those found in Ref.~\cite{Atkinson}
(corrected for the gauge covariance violating term).  We emphasize 
that in contrast to the unrenormalized equations in
Ref.~\cite{Atkinson}, for the renormalized equations this power Ansatz
is valid whether or not the bare mass $m_0$ is zero.  We demonstrate
this explicitly in Fig.~(\ref{fig:numan}), where the above asymptotic
form of the mass function (dashed line) is compared to previously
published numerical work (Ref.~\cite{qed4_hw_etal1}) with $m_0 \ne 0$ (solid
line).  As one can see there is excellent agreement in the region $p^2
>> M^2(p^2)$.  Quantitatively, one sees in Table~[1] that the
logarithmic slope $s$ calculated from Eq.~(\ref{eq: s def}) for a
variety of gauge parameters and couplings agrees, in most cases, with
the same quantity extracted from the published numerical work to
better than 1 part in $10^5$. 

We note that in the above calculations $\alpha < \alpha_{cr}$.  There
are, in general, several real solutions $s$ to Eq.~(\ref{eq: s def})
at these values of the coupling (see also Ref.~\cite{Atkinson}).
However, only one of these matches smoothly onto the perturbative
solution and it is this one which is used above.  The other solution
appears to be a spurious byproduct of the linearized approximation
(\ref{eq:bifmass}) to
Eq.~(\ref{eq:renmass}) (Solutions which do not
match smoothly onto perturbation theory can arise in
Eq.~(\ref{eq:bifmass}) if the integrals diverge like ${1 \over
\alpha}$ due to IR divergences; this cannot occur in the exact
Eq.~(\ref{eq:renmass}) because of the regulating mass in the
denominator).  Because $s$ is real there are no oscillations in the mass
function and hence there is no chiral symmetry breaking for these
couplings.  For $\alpha > \alpha_{cr}$, however, the solutions $s$
become complex. As Eq.~(\ref{eq: s def}) defines $s$ as a real analytic 
function of $\alpha$,
it is clear that if $s$ is a solution, then so is $s^*$.  In this case
the correct asymptotic tail to the oscillating, but real, mass
function (Eq.~\ref{eq:renmass}) is obtained by a suitable
superposition of these two particular solutions to
Eq.~(\ref{eq:bifmass}).  The oscillations are the indication that,
above $\alpha_{cr}$, chiral symmetry is broken in quenched QED
employing the CP vertex.

\noindent

\normalsize

%%%%%%%%%%%%%%%%%%%%%%%%%%%%%%%%%%%%%%%%%%%%%%%%%%%%%%%%%%%%%

\subsection{The anomalous mass dimension }
The analytical expression for the asymptotic behaviour of the mass function
is closely related to the anomalous mass dimension of the theory, i.e.
\begin{equation}
\gamma_m \> = \> {d \log Z_m \over d \log \mu}\> \rightarrow \> 2 s\;\;\;,
\end{equation}
where we have used $Z_m \equiv m_0/M_\mu$ and the asymptotic form
of the mass function in Eq.~(\ref{eq:bifmass}).  Hence Eq.~(\ref{eq: s def}) 
defines, within the off-shell renormalization scheme used here, 
the prediction for the asymptotic value of the anomalous mass dimension obtained 
with the CP vertex 
as a function of the coupling and the
gauge parameter $\xi$.  For convenience, let us restrict ourselves to Landau
gauge, in which case one obtains 
\be
\alpha \>= \>-\frac{8}{3}\frac{1}{\left[ 
			\cot\left (\frac{\pi \>\gamma_m}{2}\right )
                        +\frac{1}{\pi}
			 \frac{\gamma_m^2\>+\>2\>\gamma_m\>-\>12}{\gamma_m(2\>-\>\gamma_m)}
			   \right]}\;\;\;.
\label{eq:landau}
\ee

\begin{figure}[htb]
\begin{psfrags}
        \psfrag{Regamma}{\small\hspace*{-3mm}  Re $\gamma_m$\hspace*{-6mm}}
	\psfrag{Realpha}{{\small Re $\alpha$}}
	\psfrag{Imalpha}{{\small  Im $\alpha$}}
	\psfrag{Imgamma}{{\small\hspace*{-3mm} Im $\gamma_m$\hspace*{7mm}}} 
%\hspace*{1cm}
%
\centerline{
%   \parbox{8cm}{ 
	\epsfig{figure=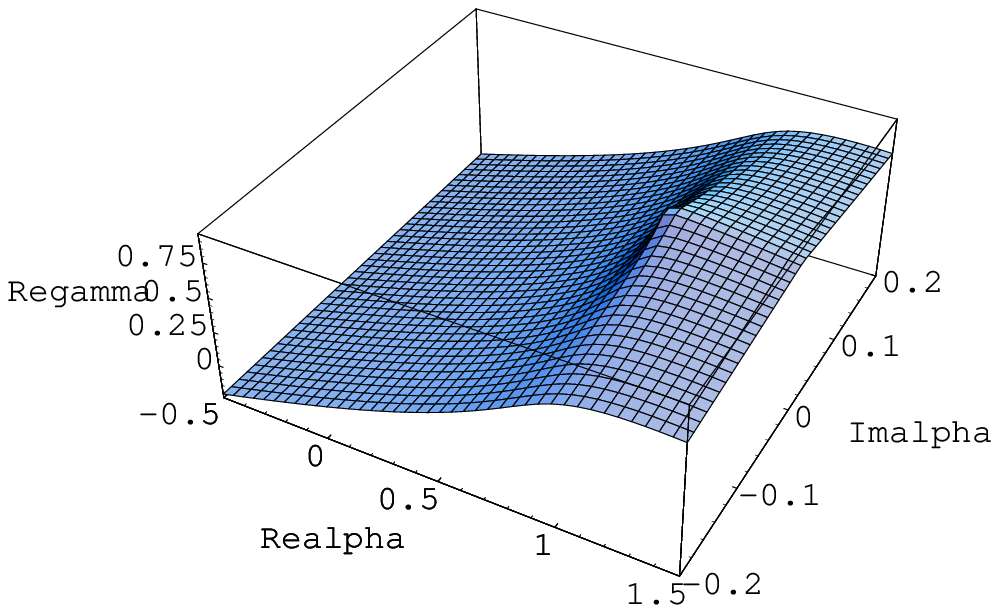,width=7cm, height=5cm} 
%}
	\hspace {0.1cm}
%   \parbox{8cm}{ 
	\epsfig{figure=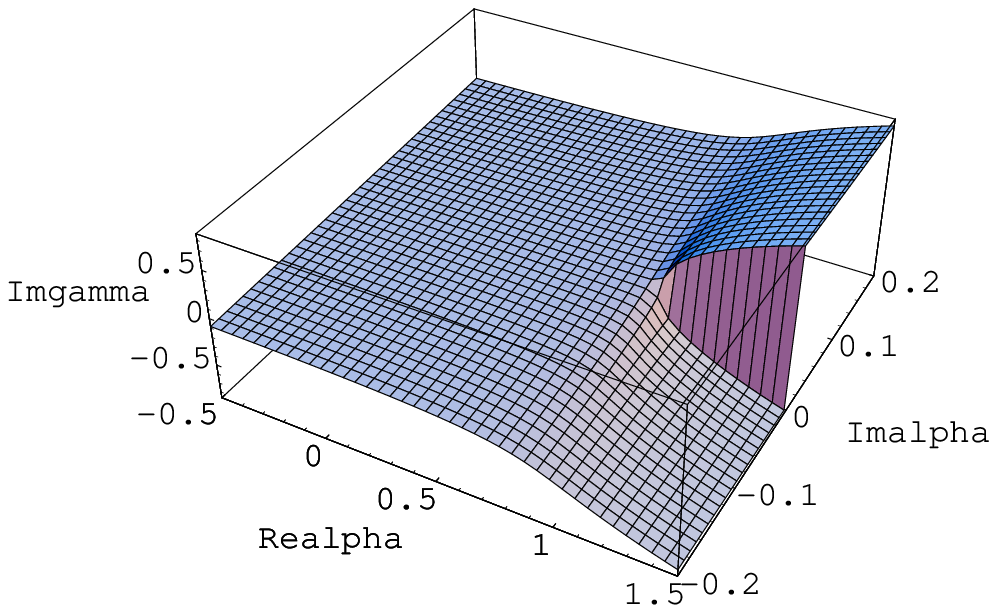,width=7cm, height=5cm}  
%}
           }
\end{psfrags}
\vspace*{5mm}
\caption{The real and imaginary part of the anomalous mass dimension versus 
complex coupling constant. }
\label{fig:complex}
\end{figure}

It is amusing to note the similarity of this equation, obtained within
the framework of Dyson-Schwinger equation studies, to that obtained
for the anomalous mass dimension within the MS scheme
using a completely different nonperturbative approach~\cite{Andreas1},
namely one based on the use of a variational principle for the evaluation
of the functional integral, i.e.
\be
\alpha \> = \> {4 \over 3} \left (1+\gamma_m^{\rm var}\right )
\cot \left ( {\pi \over 2}\>{1 \over 1 +\gamma_m^{\rm var}} \right )\;\;\;.
\label{eq: gamma var}
\ee
Not only are both equations implicit equations for $\gamma_m$, but
even the same transcendental function is involved in both.  Actually,
the similarity does not end there.  It is a simple matter to
numerically solve Eq.~(\ref{eq:landau}) for {\it complex} $\alpha$ in
order to investigate its analytic structure.  The result, for both the
imaginary and real part of the anomalous mass dimension, is shown in
Fig.~(\ref{fig:complex}). A branch point in $\alpha$ is clearly visible
on the positive real axis (the direction of the cut is arbitrary).
Its position may be determined via a bifurcation analysis of
Eq.~(\ref{eq: s def}) or, equivalently, Eq.~(\ref{eq:landau}).  Its
numerical value is $\alpha=0.933667$ (in Feynman gauge it is
$\alpha=0.890712$) and it coincides, of course, with the
well-known critical coupling associated with the CP
vertex~\cite{Atkinson,qed_dim_reg} above which the theory breaks 
chiral symmetry.  The variational result for the same quantity in the 
MS scheme, Eq.~(\ref{eq: gamma var}),
 also results in a
branch point, however this time it is at a complex value of
$\alpha$~\cite{Andreas1,Andreas2}.  Hence no sign of chiral symmetry
breaking for $\alpha > \alpha_{cr}$ was seen in $\gamma_m^{\rm var}$
in that work.

Apart from the {\it presence} of a branch point at finite $\alpha$, the
other notable feature in Fig.~(\ref{fig:complex}) (also found for
$\gamma_m^{\rm var}(\alpha)$ in Refs.~\cite{Andreas1,Andreas2}) is the
{\it absence} of a branch point, or any other nontrivial analytic
structure, at $\alpha=0$, i.e. the radius of convergence of the
perturbation expansion of $\gamma_m(\alpha)$ as obtained with the
CP vertex is finite.  This is in
marked contrast to general expectations from studies of high order
perturbation theory~\cite{HOPT}, where the factorial growth in the
number of diagrams results in a vanishing of the radius of convergence
of the perturbation expansion of a general Green function.  It is difficult to 
specify whether or
not the CP vertex correctly reproduces the factorial growth in the
number of diagrams.  What can be said with certainty, however, is that
clearly the Dyson-Schwinger equation for the fermion propagator does not
itself generate a branch point at $\alpha=0$ like it does, for
example, at $\alpha=\alpha_{cr}$.  If a branch point at $\alpha=0$ is present in
the exact theory then it already needs to be present in the vertex
itself.

Finally, associated with the above analytical structure,
one can derive the behaviour of high order perturbation theory
for $\gamma_m(\alpha)$ from Eq.~(\ref{eq:landau}), using the methods discussed
in~\cite{Andreas2}.  We define the expansion coefficients $c_n$ through
\be
\gamma_m(a)&=&\sum_{n=1}^{\infty} c_n\> a^n \quad\quad\quad a \equiv {\alpha \over \pi}
\label{eq: gamser}
\ee
so that the radius of convergence (in $a$) is given by
$ \lim_{n \rightarrow \infty} |c_n|^{-1/n}$.
The asymptotic behaviour of these coefficients may be derived most
easily by converting Eq.~(\ref{eq: s def}) into a
differential equation: 

\be
-\frac{8}{3a^2}=\frac{1}{2}\left \{
                      \pi^2+\left[-\frac{8}{3a}+1+3f(a)+g(a)\right]^2 
                \right \}{\gamma_m'(a)}
                +3f^{\prime}(a)+g^{\prime}(a)
\label{eq:differential}
\ee

\noindent
where we have defined
$f(a)=2/{\gamma_m(a)}$ and $g(a)=1/[1-\gamma_m(a)/2]$. These functions
have the 
perturbative expansions 
\be
f(a) = \frac{2}{a\,c_1} \sum_{n=0}^{\infty}b_n\>a^n \quad \quad
g(a) = \sum_{n=0}^{\infty}d_n\>a^n 
\label{eq:expansion}
\ee
where the coefficients $b_n$ and $d_n$ are related to the $c_n$'s 
through~\cite{GR}
\be 
b_0 \> = \>1\quad &\quad& d_0 \> = \> 1 \nonumber \\
b_n \> = \> - \sum_{k=1}^{n} b_{n-k} {c_{k+1} \over c_1}
\quad &\quad&
d_n \> = \>  {1 \over 2} \sum_{k=1}^{n} d_{n-k}\> c_k \quad \quad\quad n > 0
\label{eq:expansion 2}
\ee
These equations, together with Eqs.~(\ref{eq: gamser}) and~(\ref{eq:expansion}) 
substituted
into Eq.~(\ref{eq:differential}), define the expansion coefficients $c_n$ and
may be solved iteratively to very high orders.  For the first few orders
we obtain
\be
\gamma_m(\alpha)&=& \frac{3}{2}\frac{\alpha}{\pi}
              +\frac{9}{8}\left(\frac{\alpha}{\pi}\right)^2
   +\frac{9}{64}(9+\pi^2)\left(\frac{\alpha}{\pi}\right)^3 \nonumber\\
&&\hspace{5mm}
   +\frac{81}{256}(6+\pi^2)\left(\frac{\alpha}{\pi}\right)^4
   +\frac{81}{2560}(105+20\pi^2+\pi^4)\left(\frac{\alpha}{\pi}\right)^5
   +{\cal{O}}(\alpha^6)
\ee
while for large $n$ the behaviour of the coefficients approaches the form
\be 
c_n=\left({\alpha_{cr}\over \pi}\right )^{-n}\frac{e^{-\beta_0}}{n^{3/2}} 
\label{eq:cn}
\ee 
with $\beta_0 \approx 0.52$.  The absence of a term of the form $n^n$
indicates the absence of factorial growth mentioned previously and the power $3/2$ in the 
denominator is indicative of the fact that close to the branch point
the cut has the character of a square root.

%%%%\nonewpage       
\section{Conclusions and Outlook}
\label{sec_conclusions}

In this letter we have argued that there are significant advantages in
performing nonperturbative calculations based on the use of
Dyson-Schwinger equations in a manner different to that commonly used
in the literature.  The central idea is to work with renormalized
quantities only, allowing one to completely eliminate the regulator.
Within QED, this allows one to avoid spurious problems associated with
the usual cut-off regulator, such as the breaking of gauge covariance
in the fermion propagator and the introduction of unwanted quadratic
divergences into the photon propagator.  Furthermore we have shown that,
within quenched QED, it allows one to gain an analytical understanding
of both the mass function at large momenta as well as the analytic
structure of the anomalous mass dimension of this theory.  We expect
that the main advantage of the method, however, will be that one can
restrict the numerical task of solving the Dyson-Schwinger equations
of any renormalizable theory to a relatively small region of momenta.  For
example, we anticipate that
this feature will make it feasible to study full, rather than
quenched, QED$_4$.  Furthermore, it is our hope that the numerical
task  of solving a coupled set of Dyson-Schwinger equations
is sufficiently simplified by the approach discussed here to allow
one to consider pushing the truncation of this set beyond the equations
for the two-point functions.

\begin{acknowledgements}

We are grateful to F.~Hawes for providing us with the data of 
Ref.~\cite{qed4_hw_etal1} used in preparing Table 1 and T.~Sizer
for helpful conversations about this work, as well as a careful reading
of this manuscript.
\end{acknowledgements}

%=======================================================================
%          Bibliography:
%-----------------------------------------------------------------------

%\end{thebibliography}
\end{document}